\documentstyle[11pt,newpasp,twoside,psfig]{article}
\begin{document}

\title{The AGN Paradigm: Radio-Quiet Objects}
\author{Sylvain Veilleux\altaffilmark{1}}
\affil{Department of Astronomy, University of Maryland, College Park, MD 20742 USA}

\altaffiltext{1}{Cottrell Scholar of the Research Corporation}

\setcounter{page}{1}
% this page number will be filled later by the editors....
\index{Veilleux, S.}

\begin{abstract}
The current paradigm for radio-quiet AGNs is reviewed, taking into
account new results from recent large-scale surveys carried out from
the ground and from space. Topics include structure of the central
engine, AGN demography, fueling/triggering processes, and connection
between the supermassive black hole, host galaxy, circumnuclear
starburst and AGN. Dependences on AGN power and lookback time are
pointed out in the discussion.  Suggestions for future avenues of
research are mentioned in the last section.
\end{abstract}

\section{Introduction}

A quick search through the NASA/ADS Abstract Service indicates that
more than $\sim$ 5000 papers were published over the past five years
on AGNs! Clearly, a comprehensive review of this literature is well
beyond the scope of the present article. The present review highlights
important results from recent large-scale ground-based surveys and
space missions, and describes the consequences of these results on our
understanding of radio-quiet AGNs. The standard paradigm for
radio-quiet AGNs is described in \S 2 along with a few key supporting
observations. Important new constraints on this standard picture have
been derived from recent UV/optical, infrared, and X-ray surveys;
these are discussed in \S 3. In \S 4, the nature of the connection
between the supermassive black hole, host galaxy, circumnuclear
starburst, and AGN is briefly reviewed taking into account recent new
data. The issue of AGN fueling/triggering is addressed in \S 5. A
summary of the outstanding issues is given in \S 6 along with
suggestions for future avenues of research.

\section{The Standard Picture}

\begin{figure}[htb]
\centerline{
}
\vskip 2.6in
\caption{Standard Unification Model for Radio-Quiet AGNs (after a sketch of
radio-loud AGNs by Urry \& Padovani 1995).}
\label{fig1}
\end{figure}

Figure 1 is an idealized representation of the inner structure of
radio-quiet AGNs. Direct observational support for this picture comes
from the detection in a few radio-quiet AGNs of H$_2$O-masing
disk-like structures in orbit around central masses of a few $\times$
10$^6$ -- 10$^7$ M$_\odot$ on scales of 0.1 -- 1.0 pc ($\sim$ 10$^5$
$r_g$, where $r_g = GM/c^2$ is the gravitational radius). The most
convincing cases so far are NGC~4258 (Miyoshi et al. 1995), NGC~1068
(Greenhill et al. 1996; Greenhill \& Gwinn 1997; Gallimore et al.
1997, 2001), and NGC~4945 (Greenhill, Moran, \& Herrnstein 1997). A
survey by Braatz, Wilson, \& Henkel (1997) failed to detect H$_2$O
masers in Seyfert 1s, perhaps indicating a less favorable face-on disk
geometry in these objects (small $N_H$ along the line of sight), as
expected in the unification model of Seyfert galaxies
(Fig. 1). Indirect evidence for accretion disk in radio-quiet AGNs
comes from the presence of broad ($\ga$ 70,000 km s$^{-1}$ in
MCG--06-30-15), skewed and redshifted Fe K$\alpha$ lines in Seyfert
1s. This line probes the purported accretion disk within a few $r_g$
of the black hole (e.g., Tanaka et al. 1995; Iwasawa et al. 1996,
1999; Nandra et al. 1997; Guainazzi et al. 1999; Wilms et al. 2001;
Lee et al. 2002; Fabian et al. 2002; Turner et al. 2002; see
contribution by Fabian in this Volume). The presence of large-scale
(photo-)ionization cones in radio-quiet AGNs is further indirect
evidence for an inner disk structure in these objects (e.g., Haniff,
Wilson, \& Ward 1988; Pogge 1989; Tadhunter \& Tzvetanov 1989; Wilson
\& Tzvetanov 1994; Mulchaey et al. 1996a, 1996b; Pogge et al. 2000;
Veilleux 2002; Kinkhabwala et al. 2002).

Historically, spectropolarimetric studies of radio-quiet AGNs have
been the main driver behind the unification model of Seyfert galaxies
(e.g., Antonucci \& Miller 1985; Miller, Goodrich, \& Matthews 1991).
But recent spectropolarimetric surveys indicate that only $\sim$ 30 --
50\% ({\em i.e.}~not all) of Seyfert 2 galaxies harbor hidden
broad-line regions (HBLRs; e.g., Tran 2001).  Seyfert 2s with HBLRs
tend to have larger radio-to-FIR flux ratios and warmer dust
temperatures than those without (e.g., Miller \& Goodrich 1990; Kay
1994; Heisler, Lumsden, \& Bailey 1997; Moran et al. 2000; Tran
2001). These trends have been interpreted by some to imply the
existence of two separate classes of Seyfert 2s. However, several
authors have been quick to point out the importance of the
orientation-dependent selection criteria used in many of these studies
(e.g., Antonucci 2001; Alexander 2001; Lumsden et al. 2001; Gu \&
Huang 2002). Biases in luminosity, starburst contribution, and
covering factor are necessarily present in these studies, so caution
should be used when interpreting the results.

Recent results from near- and mid-infrared spectroscopy of radio-quiet
AGNs have provided further support for the unification model.
Obscured BLRs have been detected at $\sim$ 2 -- 4 $\mu$m in a number
of UV- and infrared-selected Seyfert 2 galaxies (e.g., Veilleux,
Goodrich, \& Hill 1997a; Veilleux, Sanders, \& Kim 1997b, 1999b; Lutz
et al. 2002).  Clavel et al. (2000) also note that the weaker 7-$\mu$m
continuum and larger equivalent widths of the PAH feature in Seyfert
2s relative to Seyfert 1s can be explained if the continuum in the
Seyfert 2s is more strongly extinguished ($A_V = 92 \pm 37$
mag). Studies at near- and mid-infrared wavelengths can only probe
down to $N_H$ $\approx$ few $\times$ 10$^{22}$ cm$^{-2}$, {\em i.e.}
considerably smaller than the expected column density of the dusty
torus in AGN. Hard X-ray observations have extended the range of
column densities up to $\sim$ 10$^{24}$ cm$^{-2}$. Maiolino et
al. (1998) and Risaliti, Maiolino, \& Salvati (1999) have used
BeppoSax, ASCA, and GINGA data to show that Seyfert 2 galaxies
generally have larger obscuration than Seyfert 1s, as expected in the
standard unification model. However, these observations do not put any
constraints on the exact geometry and location of the obscuring
material in Seyfert 2s (e.g., inner disk/torus {\em vs.} galaxy-scale
circumnuclear material?).

Optical, UV, and X-ray observations of AGNs indicate that the simplest
form of the standard picture (Fig. 1) is not tenable and must be
modified to account for dependences on AGN luminosity. Optical
spectroscopy of infrared-selected AGNs (in which orientation-dependent
biases are less important) indicates that the ratio of Seyfert 2s to
Seyfert 1s decreases significantly with infrared luminosity (e.g.,
Veilleux et al. 1995; Veilleux, Kim, \& Sanders 1999a).  The
well-known decrease of the C IV $\lambda$1549 equivalent widths with
increasing luminosity ({\em i.e.}  the UV Baldwin effect; e.g., Osmer
\& Shields 1999; Espey \& Andreadis 1999) is another manifestation of
this strong luminosity dependence.  The Fe K$\alpha$ line and Compton
scattering hump also are weaker in QSOs than in Seyferts (the X-ray
Baldwin effect; e.g., Nandra et al. 1997).  Finally, UV and X-ray
narrow absorption lines are rarely seen in the spectra of quasars but
not so in Seyferts (e.g., Turner et al. 1993; Mathur et al. 1999;
Nicastro et al. 1999; Kaastra et al. 2000, 2002; Kaspi et
al. 2002). All of these observations can be explained in the context
of the unification model if the opening angle, or location of the
inner edge, of the disk/torus/wind structure is allowed to vary with
luminosity (e.g., Lawrence 1991;~Hill~et~al.~1996;~Elvis~2000).

\section{New Constraints from Recent Surveys}

The discussion in this section focusses on moderate-to-high luminosity
AGNs. The reader should refer to reviews by V\'eron-Cetty \& V\'eron
2000, Ho (2002), Barth (2002) and contributions in this Volume by
Filippenko and Ho for discussions of recent results on low-luminosity
AGNs (LLAGNs). Radio-loud AGNs are reviewed by Urry in this Volume so
radio surveys are not discussed here. 
%e.g., FIRST; Becker, White, \& Helfand 1995, ApJ, 450, 559

\subsection{UV/Optical: HES, 2dF, and SDSS}

Several new surveys have improved our knowledge of AGN demography as a
function of redshift; three of them are discussed here. The results
from the Hamburg/ESO Bright QSO Survey (HES; e.g., Wisotzki et
al. 2000) indicate that the local QSO density is $\sim$ 50\% larger
than previously thought (e.g., Bright Quasar Survey of Schmidt \& Green
1983). The 2dF QSO Redshift Survey (e.g., Boyle et al. 2000) and the
Sloan Digital Sky Survey (SDSS; e.g., Fan et al. 2001) have nicely
confirmed that the QSO density rises steeply up to $z \sim$ 2 and
decreases beyond $z \sim 3$.
% (e.g., Schmidt, Schneider, \& Gunn 1995).
At the time of writing, SDSS has revealed four quasars at $z >
5.8$. This set of high-$z$ QSOs should at least double in size by the
time the survey is completed. These objects will provide important new
constraints on the epoch of QSO/galaxy formation (see, e.g., Haiman \&
Cen 2002).

\subsection{Infrared: 2MASS}

The Two Micron All Sky Survey (2MASS) has uncovered a large population
of red ($J - K > 2$) AGNs at relatively small redshifts (median $z$
$\sim$ 0.25; Cutri et al. 2001; see also Gregg et
al. 2002). Approximately 75\% of these sources are previously
unidentified AGNs whose space density ($\sim$ 0.5 deg$^{-2}$) is
comparable to that of optical/UV selected AGNs of the same IR
magnitudes. A large fraction ($\sim$ 80\%) of these objects are Type 1
AGNs. Chandra follow-up observations of some of these AGNs indicate
that all of them are X-ray faint, with the reddest being the faintest
in X-rays (Wilkes et al. 2002). Interestingly, these broad-lined AGNs
show significant absorption (several of them have $N_H \sim 10^{22}$
cm$^{-2}$) and may be important contributors to the cosmic X-ray
background (CXB; cf.~\S 3.3; see also Webster et al. 1995; Benn et
al. 1998; Whiting et al. 2001 for analyses of similar PKS
radio-selected red QSOs).

\subsection{X-Rays: BeppoSax, Chandra, and XMM-Newton}

As mentioned in \S 2, X-ray observations generally provide support for
the unification model of Seyfert galaxies. However, there are a number
of important exceptions. These mismatches in the optical -- hard X-ray
classification fall into three broad categories: (1) Several
broad-line AGNs show significant absorption in the X-rays (e.g.,
Seyferts and QSOs: Fiore et al. 2001, 2002, Wilkes et al. 2002;
BALQSOs: Mathur et al.  2001, Gallagher et al. 2002). (2) Deep X-ray
surveys have revealed a large population ($\sim$ 40 -- 60\%) of X-ray
bright sources with no obvious optical AGNs (e.g., 
%Mushotzky et al. 2000; Hornschemeier et al. 2001; 
Barger et al. 2001, 2002; Tozzi et al. 2001;
%Hasinger et al. 2001; 
Stern et al. 2002b; Fiore et al. 2002). (3) A number of Seyfert 2
galaxies show no obvious X-ray absorption (e.g., Ptak et al. 1996;
Bassani et al. 1999; Pappa et al. 2001; Panessa \& Bassani 2002). The
first category of objects may be explained in the context of the
unification model if dust near the AGN is different from Galactic
(e.g., different dust/gas ratio, metallicity, grain size; Veilleux et
al. 1997a; Maiolino et al. 2001) or is not co-spatial with the
(neutral and ionized) material producing the X-ray absorption (e.g.,
Veilleux et al. 1997a). X-ray bright sources with no obvious optical
AGNs may be powered through inefficient ADAF-like accretion with
$L/L_{\rm Edd} < 10^{-3}$, or perhaps the apparent lack of optical
AGNs in these sources is simply due to strong dilution effects by the
host galaxy light (e.g., Moran, Filippenko, \& Chornock
2002). Finally, narrow-line AGNs with no obvious X-ray absorption have
often been shown to be (nearly) Compton thick when observed in the
hard X-rays. Half of all nearby optically selected Seyfert 2s may fall
in this category [e.g., Risaliti et al. 1999; well-known examples
include NGC~1068, NGC~4945 (Done, Madejski, \& Smith 1996), and
Circinus (Matt et al. 1999)].  Luminous examples also exist, although
they may be less common: e.g., QSO-2s (Norman et al.  2002; Crawford
et al. 2002; Stern et al. 2002a), X-ray detected Extremely Red Objects
(EROs) with $R - K > 5$ (e.g., Alexander et al. 2002; Mainieri et
al. 2002),
%Brusa et al. 2002; 
and ultraluminous infrared galaxies (e.g., NGC~6240; Iwasawa \&
Comastri 1998; Vignati et al. 1999).
%Mrk 231; Maloney \& Reynolds 2000

The improved sensitivity of the current X-ray facilities has allowed
to search for possible evolutionary effects in the X-ray spectra of
quasars and AGNs over a broad redshift range. Brandt et al. (2002) and
Mathur, Wilkes, \& Ghosh (2002) recently failed to find convincing
evidence for a redshift dependence of the optical -- X-ray slope in
QSOs out to $z \sim 6$. Bechtold et al. (2002) come to a different
conclusion using a larger comparison sample. A significant population
of buried Type 2 AGNs with peak emissivity around $z \sim 0.8$ (rather
than $z \sim 2 - 3$ for Type 1 AGNs; \S 3.1) appears to be needed to
explain the properties of the CXB and the redshift dependence of the
number density of X-ray sources in the deep Chandra and XMM surveys
(e.g., Rosati et al. 2002; Franceschini, Braito, \& Fadda 2002). These
results may imply that Type 1 and 2 AGNs follow different evolutionary
paths, in disagreement with the AGN unification model. Note, however,
that the recent discovery of a large population of strongly absorbed
Type 1 AGNs by Wilkes et al.  (2002) may affect these conclusions.

\section{Host Galaxy -- SMBH -- Starburst -- AGN Connection}

Ground-based and $HST$ observations of the stellar and gas kinematics
near the center of ``normal'' (inactive) galaxies have revealed a
close connection between the mass of the supermassive black hole at
the center of each of these objects and the mass of the spheroidal
component (e.g., Kormendy \& Richstone 1995; Faber et al. 1997;
Magorrian et al. 1998; Gebhardt et al. 2000; Kormendy \& Gebhardt
2001; Merritt \& Ferrarese 2001; Tremaine et al. 2002) and perhaps
also the mass of the dark matter halo (Ferrarese 2002).  The results
of reverberation mapping in AGNs suggest that the $M_{\rm BH} - M_{\rm
bulge} - \sigma$ relations found in inactive galaxies also apply to
active galaxies (e.g., Wandel, Peterson, \& Malkan 1999; Kaspi et
al. 2000; Peterson \& Wandel 2000; Laor 2001; Onken \& Peterson 2001;
McLure \& Dunlop 2002; Ferrarese et al. 2001; 
%Woo \& Urry 2002;
although see Krolik 2001).  These results suggest a causal connection
between spheroid/galaxy formation, black hole growth and AGN activity
(e.g., Efstathiou \& Rees 1988; Small \& Blandford 1992; Haehnelt \&
Rees 1993; Chokshi 1997; Silk \& Rees 1998; Haiman \& Loeb 1998;
Fabian 1999; Kauffmann \& Haehnelt 2000; Burkert \& Silk 2001; Adams,
Graff, \& Richstone 2001). This tight SMBH -- host galaxy connection
is discussed in more detail by Urry, Peterson, and Merritt in this
Volume.

A strong connection also exists between starbursts and AGNs (Veilleux
2001; also Ward this Volume). Nuclear starbursts appear to be present
in several Seyfert galaxies, based on the strength of UV and optical
absorption and emission features from young/intermediate-age stars in
the nuclear spectra of these objects (e.g., Terlevich, Diaz, \&
Terlevich 1990; Heckman et al. 1997; Gonzalez Delgado et al. 1998,
2001). The presence of extended soft thermal X-ray emission in
Seyferts also supports this scenario (e.g., Levenson, Weaver, \&
Heckman 2001). Claims that starbursts are more common in Seyfert 2s
than in Seyfert 1s have been made for many years (e.g.,
Rodriguez-Espinosa, Rudy, \& Jones 1986; Pier \& Krolik 1993; Maiolino
et al. 1995; Nelson \& Whittle 1996; Gonzalez Delgado et al. 2001; Gu,
Dultzin-Hacyan, \& de Diego 2001), but orientation-dependent selection
criteria may seriously bias some of these samples and cause the
apparent Seyfert 1 / Seyfert 2 dichotomy.

The nuclear starbursts in active galaxies may have a very strong
impact on the evolution of the host galaxy and the AGN itself. The
stellar winds and supernovae associated with the nuclear starburst may
deposit sufficient amounts of mechanical energy at the centers of
these objects to severely disrupt the gas phase of these systems and
result in large-scale galactic winds. A well-known example of
starburst-driven wind in an active galaxy is NGC~3079. Detailed
optical, radio, and X-ray studies of this object have revealed the
presence of a powerful galactic wind that is strongly interacting with
the ambient material of the host galaxy (e.g., Veilleux et al. 1994;
Cecil et al. 2001; Cecil, Bland-Hawthorn, \& Veilleux 2002). The wind
event is clearly disturbing the distribution of gas within the central
kpc of this object and may therefore also affect the feeding of the AGN.
The frequency of starburst-driven winds in AGNs is poorly constrained
(see Veilleux \& Rupke 2002 for a discussion of a promising search
technique), but if common these winds may provide negative feedback
on the fueling mechanisms of AGNs (e.g., Rupke, Veilleux, \& Sanders
2002).

\section{Fueling and Triggering Mechanisms}

The broad range in luminosity of AGN ($<$ 10$^9$ -- 10$^{14}$
L$_\odot$) implies accretion rates of order $\la$ 0.001 -- 100
M$_\odot$ yr$^{-1}$ (assuming a radiative efficiency of 10\% in rest
mass units); the required accretion rates are very small for LLAGNs
but quite substantial for QSOs.  The requirements on the
fueling/triggering processes for AGNs are therefore highly dependent
on the AGN luminosity. Local processes are sufficient to power LLAGNs
and Seyferts, but galactic-scale phenomena may be required to explain
powerful QSOs. A broad range of processes including galaxy
interactions and mergers, large-scale and nuclear bars, and nuclear
gaseous spirals have been proposed to account for the fueling of
Seyferts and LLAGNs (see Combes 2000 and the contribution by Combes in
this Volume). The lack of obvious excess of companions and mergers
seems to rule out the possibility that galaxy interactions and mergers
are solely responsible for triggering these objects (e.g.,
Fuentes-Williams \& Stocke 1988; Dultzin-Hacyan et al. 1999; De
Robertis, Yee, \& Hayhoe 1998; Virani, De Robertis, \& VanDalfsen
2000). A slight (2.5-$\sigma$) excess of bars appear to be present
among Seyferts (e.g., Knapen, Shlosman, \& Peletier 2000; Laine et
al. 2002; Knapen this Volume; although see Mulchaey \& Regan 1997;
Regan \& Mulchaey 1999; Martini et al. 2001), but this result cannot
explain Seyfert activity in galaxies without bars. Nuclear (0.1 -- 1
kpc) gaseous spirals have been detected in most Seyferts but they also
appear to be present in several normal galaxies and therefore cannot
be the only reason for the Seyfert activity (e.g., Ford et al. 1994;
Dopita et al. 1997; Laine et al. 1999, 2001; Martini \& Pogge 1999;
Regan \& Mulchaey 1999; Pogge \& Martini 2002; Emsellen this
Volume). Most likely, these various processes help replenish a
reservoir of fuel on scales of $\sim$ 100 pc or larger, but other
processes are needed to carry the material down to sub-pc scales
(e.g., gas instabilities, stellar ejecta, magnetic
fields). Unfortunately, very little is known observationally on scales
$\la$ 10 pc.

The origin of the activity in high-luminosity objects is perhaps less
ambiguous than in low-luminosity objects. Hosts of luminous QSOs
generally appear to be elliptical galaxies (e.g., Dunlop et al. 2002;
although see McLeod \& McLeod 2001), but signs of interaction are seen
in several QSOs, especially in those with infrared excess (e.g.,
Surace, Sanders, \& Evans 2001; Canalizo \& Stockton 2000,
2001). Several of these objects contain large quantities of molecular
gas (e.g., Evans et al. 2001), suggestive of a gas-rich merger
origin. Ultraluminous infrared galaxies (ULIGs) have long been
suspected to be the progenitors of optical quasars based on the fact
that nearly all ULIGs show signs of recent galaxy mergers and
starburst/AGN activity, but only recently has there been a large
enough homogeneous sample of ULIGs to look carefully at this
question. Recent optical, near-infrared, and mid-infrared
spectroscopic surveys of the 1-Jy sample of 118 ULIGs indicate that
$\sim$ 30\% (50\%) of the objects with log[L$_{\rm IR}$/L$_\odot$]
$\ge$ 12.0 (12.3) are powered predominantly by a quasar rather than a
starburst (e.g., Kim, Veilleux, \& Sanders 1998; Genzel et al. 1998;
Veilleux, Kim, \& Veilleux 1999a; Lutz, Veilleux, \& Genzel 1999). A
recent morphological analysis of the 1-Jy sample indicates strong
trends between infrared luminosity and colors, quasar activity, and
merger phase (Veilleux, Kim, \& Sanders 2002). All ULIGs in the 1-Jy
sample are found to be in the pre-merger or final merger phase. About
$\sim$ 70\% of the extreme ULIGs with log[L$_{\rm IR}$/L$_\odot$] $>$
12.5 are single-nucleus advanced mergers. All (most) Seyfert 1s (2s)
in the sample are also advanced mergers.  The $R$- and $K^\prime$-band
profiles in 73\% of the single-nucleus advanced mergers are well
fitted over $R$ = 4 -- 12 kpc by an elliptical-like $R^{1/4}$ law (see
also Scoville et al. 2000; Cui et al. 2001). These elliptical-like
hosts follow the same $R$-band $\mu_e$ -- $r_e$ relation as normal
ellipticals, suggesting that these objects may eventually become
intermediate-luminosity (1 -- 3 L$^\ast$) elliptical galaxies if they
get rid of their excess gas or transform the gas into stars. The hosts
of ULIGs show a broad range in luminosity (mean $\sim$ 2 L$^\ast$)
which overlaps with that of QSO hosts. The $R$ -- $K^\prime$ colors of
ULIG hosts (mean $\sim$ 3.0) are also similar to those of QSO hosts
and normal ellipticals. The average half-light radius of ULIGs is 4.8
$\pm$ 1.4 kpc at $R$ and 3.5 $\pm$ 1.4 kpc at $K^\prime$, similar to
the QSO host sizes measured at $H$ by McLeod \& McLeod (2001) but
slightly smaller than those measured at $R$ by Dunlop et al. (2002).
The reason for this apparent discrepancy between the two QSO datasets
is not known. Overall, these results support the scenario in which the
ULIG is the result of the merger of two $\sim$ L$^\ast$ disk galaxies
which eventually evolves into an elliptical-like galaxy with a
powerful AGN. There are obviously exceptions to this scenario and this
is why a large sample of objects like the 1-Jy sample is needed to
draw statistically meaningful conclusions.

\section{Summary and Unanswered Questions}

The following is a list of outstanding issues and unanswered questions:
\begin{itemize}

\item[1.] There is strong general support for the AGN unification
model (Fig. 1), but there are important exceptions: (a) not all
Seyfert 2s show HBLRs. This may be a real effect or it could be due to
instrument sensitivity ({\em i.e.} not enough ``mirrors'' to scatter
the BLR emission back into our line of sight). (b) Optical and X-ray
classifications are not always consistent with each other. These
classification mismatches may be explained by non-standard dust near
AGNs, inefficient ADAF-like accretion, dilution effects by galaxy
light, or Compton thickness.

\item[2.] The luminosity dependence of the Type 2 / Type 1 AGN ratio
and the UV and X-ray Baldwin effects of the emission and absorption
lines in AGNs require that the geometry of the disk/torus/wind structure 
depends on AGN luminosity. The exact dependence relies on knowing the
fraction of obscured Type 2 AGNs, which is still subject to large
uncertainties.

\item[3.] Recent Chandra investigations of high-redshift quasars come
to conflicting conclusions regarding possible evolutionary effects of
the sources of energy and absorption.  The situation at low redshift
is less ambiguous: The properties of the CXB and results from deep
Chandra and XMM surveys appear to require a significant population of
buried Type 2 AGNs with peak emissivity around $z \sim 0.8$, rather
than $z \sim 2 - 3$ for unobscured Type 1 AGNs. These results may
imply that Type 1 and 2 AGNs follow different evolutionary paths, in
contradiction with the AGN unification model. The recent discovery of
a large population of partially obscured Type 1 AGNs at low redshifts
may affect these conclusions.

\item[4.] There is now strong evidence that the host galaxy -- SMBH
connection originally found in inactive galaxies also applies to
active galaxies. Emission-line methods to test this connection at high
redshifts are promising but will undoubtedly be less accurate (e.g.,
Nelson 2000; McLure \& Jarvis 2002; Vestergaard 2002).

\item[5.] There appears to be a symbiotic relation between starbursts
and AGNs.  The origin of this relation is not known. The surveys
suggesting a greater occurrence of (circumnuclear) starbursts in
Seyfert 2s than in Seyfert 1s are often strongly affected by selection
biases. Feedback from starburst-driven galactic winds may be one of
many side effects of this tight starburst -- AGN connection. This wind
phenomenon is particularly important in understanding AGN activity at
high redshift.

\item[6.] Minor mergers, bars, and nuclear spirals may all combine to
bring the fuel down to $\sim$ 100 pc in Seyferts and LLAGNs, but it is
not clear what happens next. There is now strong evidence that
gas-rich mergers are able to trigger some QSOs at low redshifts after
undergoing a ULIG phase, but there is very few observational
constraints on the formation process for the bulk of QSOs that were
formed at $z \sim 2 - 4$.

\end{itemize}

\acknowledgements The author congratulates the conference organizers
for a very successful meeting. Some of the work discussed in this
review was done in collaboration with D.-C. Kim, D. B. Sanders,
J. Bland-Hawthorn, G. Cecil, P. L. Shopbell, and D. S. Rupke.  This
work was supported by NASA/LTSA grant NAG 56547, NSF/CAREER grant
AST-9874973, and a Cottrell Scholarship awarded by the Research
Corporation.

%\tiny

\end{document}